\documentclass[acus]{JAC2000}

\usepackage{graphicx}

\begin{document}
\title{\flushright{T07}\\[15pt] \centering MEASUREMENT OF THE
HADRONIC CROSS SECTION AT KLOE USING THE RADIATIVE RETURN}



\author {The KLOE collaboration\footnotemark, presented by\\
Achim Denig, LNF - INFN, Frascati, Italy }



\maketitle

\renewcommand{\thefootnote}{\fnsymbol{footnote}}
\footnotetext{$^*$The KLOE collaboration: 
A.~Aloisio, F.~Ambrosino, A.~Antonelli, 
M.~Antonelli, C.~Bacci, G.~Barbiellini, F.~Bellini,
G.~Bencivenni, S.~Bertolucci, C.~Bini
% ...
}

\renewcommand{\thefootnote}{\arabic{footnote}}

\begin{abstract}
We report on the measurement of the hadronic cross section
below $1$ GeV at the electron-positron-collider DA$\Phi$NE,
using the multiple purpose detector KLOE. The radiative return,
which is due to initial state radiation
($e^+ e^- \to \gamma + hadrons$), allows us to obtain the 
cross section for {\it variable} center-of-mass-energies 
of the hadronic system from the $2m_\pi$ threshold
up to $1.02$ GeV. This measurement can be performed while
DA$\Phi$NE is running with a {\it fixed} accelerator energy on the
$\phi$ mass ($1.02$ GeV). For the exclusive process
$e^+ e^- \rightarrow \pi^+ \pi^- \gamma$, the status of the analysis
and first preliminary results of the invariant mass spectrum of 
the two-pion-state are presented. 

\end{abstract}

\section{HADRONIC CROSS SECTION AT DA$\Phi$NE}

\subsection{Motivation}

The measurement of the hadronic cross section at low energies
is of great importance for the improvement of the theoretical 
error of the anomalous magnetic moment of the muon, 
$a_\mu = (g_\mu-2)/2$.
The hadronic contribution $a_\mu^{hadr}$
is given by the hadronic vacuum polarization and 
cannot be calculated at low energies in the framework of 
perturbative QCD. Following a phenomenological approach, 
the hadronic contribution can however be evaluated 
from the measurement of R through a dispersion relation.

\begin{equation}
a_{\mu}^{hadr} = (\frac{\alpha m_{\mu}}{3 \pi})^2 \int_{4
m_{\pi}^2}^{\infty} ds \frac{R(s) \hat{K}(s)}{s^2}   ,\\
\label{amu}
\end{equation}
where $R(s)=\frac{\sigma(e^+ e^- \rightarrow
hadrons)}{\frac{4\pi\alpha^2(s)}{3s}}$ and  the kernel 
$\hat{K}(s)$ is a smooth bounded function growing from
$0.63$ at threshold to $1$ at $\infty$. Due to the $1/s^2$ dependence in
the integral, hadronic data at low energies are
strongly enhanced in the contribution to $a_\mu^{hadr}$.
The error of the hadronic contribution is therefore 
given by the limited knowledge of hadronic cross section data.
This error is the dominating contribution to the total error of
$a_\mu^{theo}$ ($\delta a_\mu^{theo} \approx \delta a_\mu^{hadr}$).
\\
We refer to Ref. e.g. \cite{eidjeg}, \cite{davhoe}, 
\cite{marc} for a detailed discussion of the 
subject and the interpretation of the actual discrepancy between
the theoretical and the experimental value for $a_\mu$: 
$a_\mu^{theo} = (11659159.7 \pm 6.7)$x$10^{-10}$ \cite{davhoe},
$a_\mu^{exp} =  (11659202.0 \pm 15.0)$x$10^{-10}$ 
(world average including E821 
measurement\footnote{The final goal of the E821 
collaboration is a measurement of
$a_\mu$ with a precision of ca. $4$x$10^{-10}$} \cite{e821}). 
In the value shown for $a_\mu^{theo}$, $\tau$ decays have been
included for the evaluation of the dispersion integral 
under the assumption of conserved vector current and isospin symmetry.
The value for $a_\mu^{hadr}$ under these assumptions is:
$a_\mu^{hadr}=(692.4 \pm 6.2)$x$10^{-10}$ \cite{davhoe}.
An updated analysis \cite{jeg}, which includes $e^+ e^-$  
{\it data only}, finds $a_\mu^{hadr} = (697.4 \pm 10.5)$x$10^{-10}$, 
where the error can be reduced to $\approx 6$x$10^{-10}$ 
if the hadronic cross 
section is measured with an accuracy of $\approx 1\%$ 
in the energy range below $1GeV$. 

\subsection{Radiative Return}

We present in this paper a complementary approach for the 
measurement of the hadronic cross section, which uses the 
radiative process $e^+ e^- \to hadrons + \gamma$, where the
photon has been radiated by one of the initial electrons
or positrons (Initial State Radiation, ISR)
\cite{spagn}. The DA$\Phi$NE 
collider is operating with a fixed center-of-mass-energy 
on the $\phi$ resonance\footnote{An energy scan at DA$\Phi$NE 
requires a non trivial modification of the interaction region 
which has been designed especially for the $\phi$ mass region.}.
Hence, the hadronic cross section in the energy range 
$(2m_\pi)^2 < Q^2 < (m_\phi)^2$ is accessible by  
{\it radiative return} ($Q^2$ is the invariant mass of the
hadronic system). In order  
to deduce the differential cross section 
$d\sigma(e^+ e^- \to hadrons)/dQ^2$ from the measurement 
$d\sigma(e^+ e^- \to hadrons + \gamma)/dQ^2$, 
a precise theoretical understanding of the 
initial state radiation process (radiation function H)
is mandatory:

\begin{equation}
Q^2 \cdot \frac{d\sigma_{hadrons+\gamma}}{dQ^2} =
\sigma_{hadrons} \cdot H(Q^2,\Theta_i)\\
\label{H}
\end{equation}

The knowledge of the function $H(Q^2,\Theta_i)$ 
(which depends on $Q^2$ and the experimental acceptance cuts $\Theta_i$)
at a precision better than $1\%$ is a challenging task. 
However, radiative corrections were computed by different groups 
up to next-to-leading-order for the exclusive hadronic 
state $\pi^+ \pi^-$  
\cite{khoze}, \cite{binner}, \cite{german}, \cite{graz}. 
We concentrated in our analysis on this important state 
($e^+ e^- \rightarrow \rho \gamma \rightarrow \pi^+ \pi^- \gamma$)
since it is the dominating hadronic reaction below 
the $\phi$ mass and the respective process 
$e^+ e^- \to  \rho \to \pi^+ \pi^-$ accounts for $62\%$ of
the hadronic contribution to $a_\mu$ (see formula (\ref{amu})). 
\\
\\
We would like to stress that the radiative return method - 
as presented here - has the merit against the conventional 
energy scan, that the systematics of the measurement
(e.g. normalization, beam energy) have to be taken into
account only {\it once} while for the energy scan they 
have to be known for {\it each} energy step. 

\subsection{Suppression of FSR}

An important issue for the radiative return method is the 
suppression of events, where the photon has been emitted
by one of the pions (Final State Radiation, FSR). 
The choice of a phase space region, where FSR is low,
can reduce this kind of background to
an acceptable limit. 
We found from Monte Carlo studies \cite{ich}, that cutting
on $E_\gamma$ and $\Theta_\gamma$ (energy and polar angle
of the photon) effectively suppresses FSR, while  
the ISR cross section remains high. ISR events are strongly 
peaked at small angles $\Theta_\gamma$, 
while FSR events essentially follow the
$sin^2\Theta_\pi$ distribution of the pion tracks. It is therefore
essential to measure $\pi^+ \pi^- \gamma$ events with an upper 
acceptance cut for $\Theta_\gamma$ at small angles. 
A cut on $E_\gamma$ 
additionally suppresses FSR due to the fact that the decay
via the $\rho$ resonance (i.e. ISR) leads to an 
enhancement of the photon energy spectrum at $\approx 220 MeV$
which is not visible in the case of FSR. 
For the following acceptance cuts - which are the ones used in our 
analysis - we find that FSR is suppressed below $ 1\%$:
\begin{eqnarray}
 &&5^o < \Theta_\gamma < 21^o ,  159^o < \Theta_\gamma < 175^o 
 \label{A1} \\ 
 &&E_\gamma > 10 MeV \label{A2} \\
 &&55^o < \Theta_\pi < 125^o \label{A3} \\ 
 &&p_\pi^T > 200 MeV (transv. momentum)  \label{A4} 
\end{eqnarray}

The effective cross section with these acceptance cuts is $4.2 nb$.
\begin{figure}[h]
\centering
\includegraphics*[width=55mm]{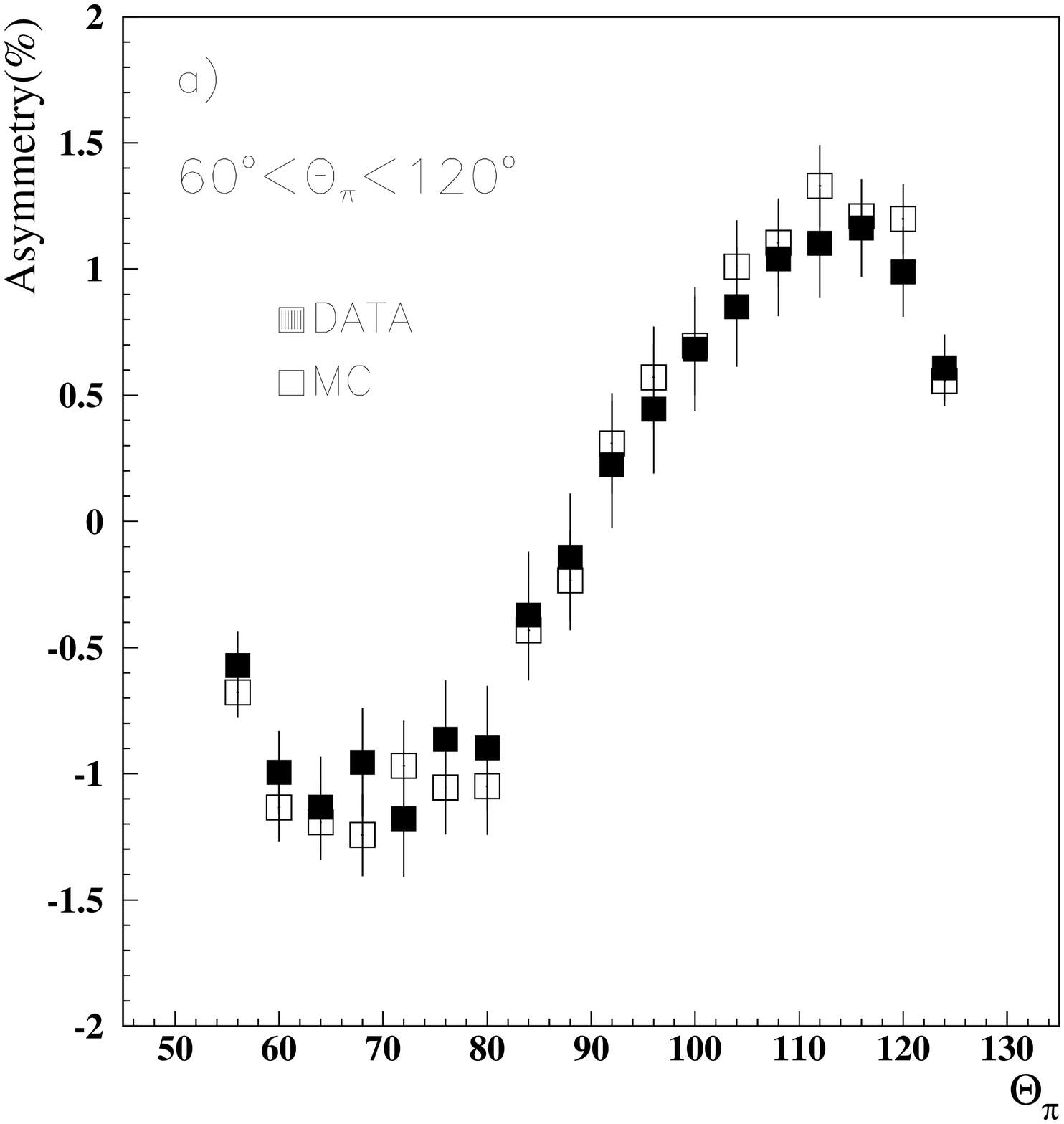}
\includegraphics*[width=55mm]{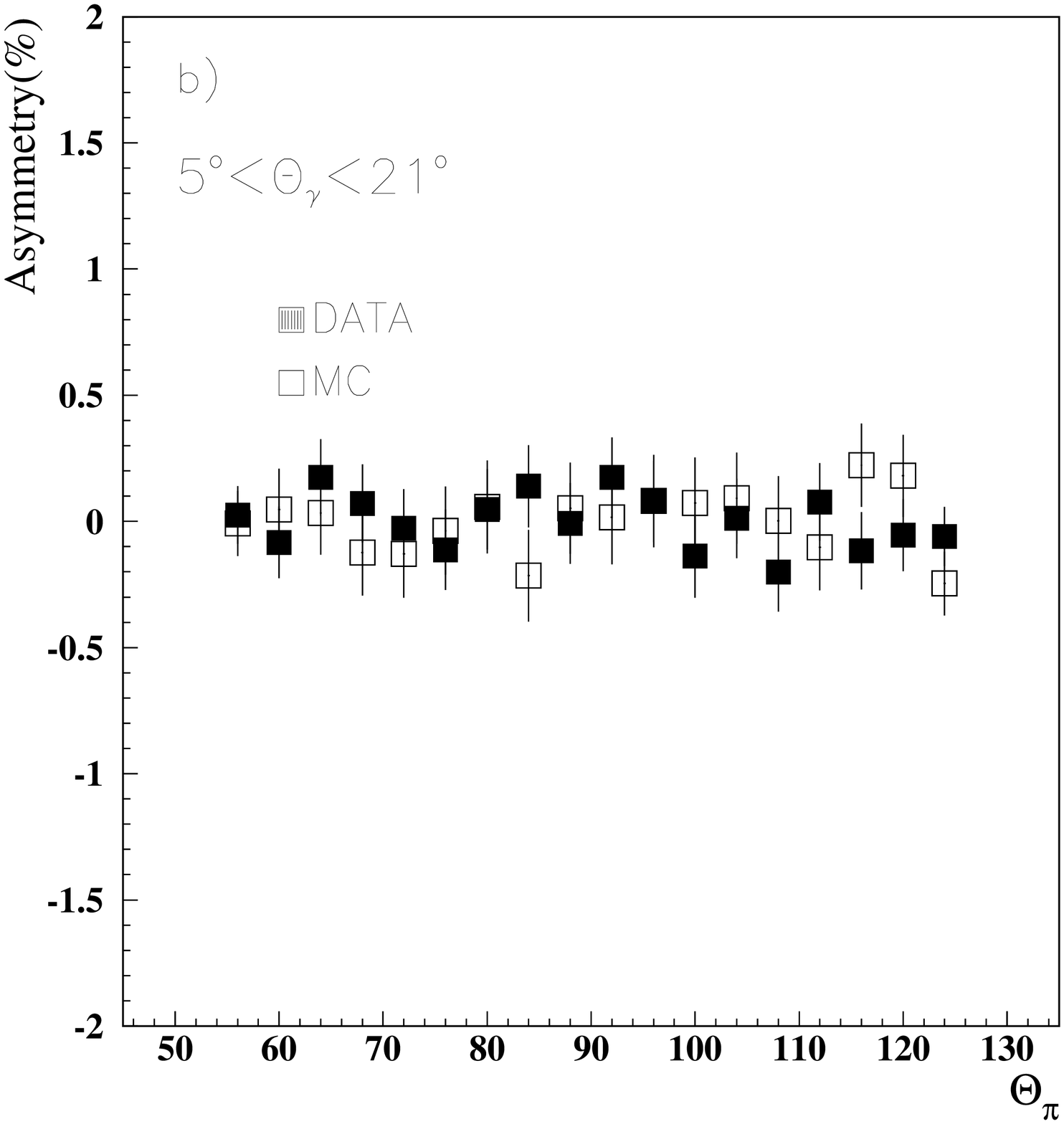}
\caption{The charge asymmetry (formula (\ref{qassym}))
is shown for 2 angular regions of the photon polar angle:
(a) $60^o < \Theta_\gamma < 120^o$, where we see a sizable
effect of the charge asymmetry due to larger FSR and 
(b) $5^o < \Theta_\gamma < 21^o$, where FSR and hence the
charge asymmetry are small.} 
\label{asymmetry}
\end{figure}
The description of FSR is model dependent and 
in the actual version of Ref. \cite{binner} a point 
like pion is assumed. The model dependence can be tested for 
$\pi^+\pi^-\gamma$ events by looking at the charge asymmetry 
of the produced pions:
\begin{equation}
A(\Theta_i) = 
\frac{N^{\pi^+}(\Theta_i)-N^{\pi^-}(\Theta_i)}
     {N^{\pi^+}(\Theta_i)+N^{\pi^-}(\Theta_i)}
\label{qassym}
\end{equation}

This charge asymmetry 
arises from the interference between ISR and FSR and is 
therefore linear in the FSR amplitude. 
We measured the charge asymmetry for 
(a) large photon angles ($60^o < \Theta_\gamma < 120^o$), 
where FSR is assumed to be large, and (b) for small 
photon angles ($5^o < \Theta_\gamma < 21^o$).
The results are illustrated in figure (\ref{asymmetry}) and show a 
very good agreement between data and Monte Carlo,
indicating that the point like model 
describes well the process of FSR within the error bars.

\begin{figure*}[t]
\centering
\includegraphics*[width=70mm]{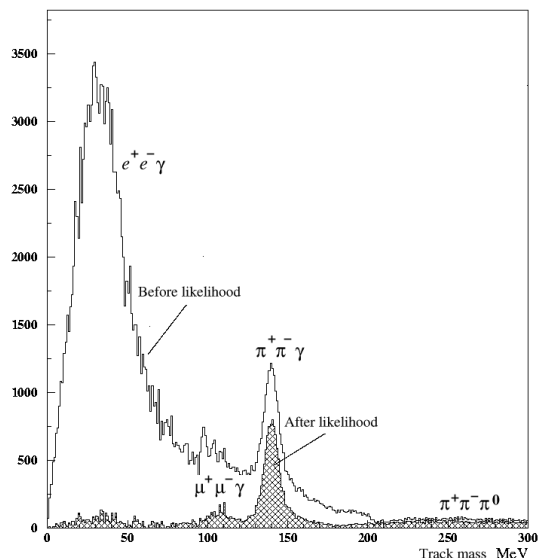}
\caption{A likelihood method has been developed to separate 
electrons from pions. The kinematic variable
track mass before and after the application of this method is shown. 
$\pi^+ \pi^- \gamma$ events are peaked at $M_{Track} = m_\pi$,
radiative Bhabha events at much smaller values. 
Other background channels are also visible 
($\mu^+ \mu^- \gamma$ and $\pi^+\pi^-\pi^0$).} 
\label{likeli}
\end{figure*}

\section{EVENT SELECTION}

In this chapter we present the event selection
for the measurement of the $\pi^+ \pi^- \gamma$ final state. 
After a very brief description of the KLOE detector, 
the selection algorithm for this signal is presented. 

\subsection{The KLOE detector}

KLOE \cite{kloe} is a typical $e^+ e^-$ multiple purpose detector 
with cylindrical geometry,
consisting of a large helium based {\it drift chamber} (DC,
\cite{dc}),
surrounded by an {\it  electromagnetic calorimeter}
(EmC, \cite{emc}) and a superconducting magnet ($B=0.6$ T). 
The detector has been
designed for the measurement of CP violation in the neutral kaon system,
i.e. for a precise detection of the decay products of $K_S$ and $K_L$.
These are low momenta charged tracks 
($\pi^\pm, \mu^\pm, e^\pm$ with a momentum range from $150 MeV/c$ 
to $270 MeV/c$) and low energy photons (down to $20 MeV$). 
\\
The DC dimensions ($3.3$ m length, $2$ m radius), 
the drift cell shapes ($2$x$2 cm^2$ cells for the inner $12$ 
layers, $3$x$3 cm^2$ cells for the outer $46$ layers) and the choice of 
the gas mixture ($90\%$ Helium, $10\%$ Isobutane; $X_0 = 900m$) 
had to be optimized for the requirements prevailing at  
a $\phi$ factory. The KLOE design results in a very good momentum 
resolution: $\sigma_{p_{\bot}} / p_{\bot} \leq 0.3\%$ at high
tracking efficiencies ($> 99\%$).
\\
The EmC is made of a matrix of scintillating fibres embedded 
in lead, which guarantees a good energy resolution 
$\sigma_E / E = 5.7 \% / \sqrt{E(GeV)}$ and excellent 
timing resolution $\sigma_t = 57 ps / \sqrt{E(GeV)} \oplus 50 ps$.
The EmC consists of a barrel and two endcaps  
which are surrounding the cylindrical DC; this gives a 
hermetic coverage of the solid angle ($98\%$). 
However, the acceptance of the EmC below $\approx 20^o$ is 
reduced due to the presence of quadrupole magnets 
close to the interaction point and does not allow us to 
measure e.g. the photon of $\pi^+ \pi^- \gamma$ events with low 
$\Theta_\gamma$ angles (as required for FSR suppression). 
\\
\\
It will be shown in the following, that an efficient
selection of the $\pi^+ \pi^- \gamma$
signal is possible, without requiring 
an explicit photon detection. 
The relatively simple signature of the signal
(2 high momentum tracks from the interaction point) and  
the good momentum resolution of the KLOE tracking detector
allow us to perform such a selection. 

\subsection{Selection Algorithm}

The $\pi^+ \pi^- \gamma$ events are selected using the following
4 steps. The selection is based on the measurement of the charged 
pion tracks by the DC, while the photon is not 
required to be detected in the EmC. 
Calorimeter information
is however used for the $\pi/e$-separation (likelihood method).  
\\
\\
i) Charged vertex in DC
\\
We require 1 vertex in the DC with 2 associated charged tracks 
close to the interaction point: $\sqrt{(x_V^2 + y_V^2)} \le 
8 cm$ and $| z_V | \le 15 cm$.
\\
\\
ii) Likelihood Method for $\pi / e$ - Separation 
\\
A fraction of radiative Bhabha events $e^+ e^- \gamma$ enters the 
kinematical selection (see next selection cut), 
giving rise to a non negligible background. In order to reject those 
events, a likelihood method has been worked out for an effective
$\pi/e$-separation. The method is based on 
the shape and energy deposition of the 
EmC clusters produced by the charged tracks 
and has been developed using independent control samples for the 
pion information ($\pi^+\pi^-\pi^0$ events) and for the
electron information ($e^+ e^- \gamma$ events).  
$98\%$ of all $\pi^+ \pi^- \gamma$ events are selected if at
least one of the two tracks has been identified as a pion by
the likelihood method. In figure (\ref{likeli}) 
the effect of the likelihood method is demonstrated in the track mass 
($M_{Track}$) distribution\footnote{The $M_{Track}$ variable 
is obtained solving the 4-momentum-conservation and resolving 
for the particle mass. }. $\pi^+ \pi^- \gamma$ events
are peaked at $M_{Track} = m_\pi$, radiative
Bhabha events at smaller values. 
\\
\\
iii) Kinematic Cut: $130.2 MeV < M_{Track} < 149.0 MeV$
\\
We perform a $\pm 9.6MeV$ ($\pm 2\sigma$)
cut on the kinematical variable $M_{Track}$, which is peaked 
at $M_{Track} = m_\pi$ for $\pi^+ \pi^- \gamma$ events 
(see again figure (\ref{likeli})).  Background of  
$\pi^+\pi^-\pi^0$ events (with $M_{Track}$ mostly $>150MeV$), 
$\mu^+ \mu^- \gamma$ (with $M_{Track}$ peaked at $m_\mu$) and
the bulk part of radiative Bhabha events $e^+ e^- \gamma$ 
($M_{Track}< 100 MeV$) are mostly rejected 
(see subchapter 3.3 on background). 
\\
\\
iv) Acceptance Cuts
\\
The missing momentum of the 2 accepted charged tracks is 
calculated and associated with the photon under the 
assumption of a $\pi^+ \pi^- \gamma$ event:
$\vec{p}_{\gamma}^{DC}  =  \vec{p_{\phi}} - \vec{p_+} - \vec{p_-}$
\footnote{$\vec{p_{\phi}}$ is the $\phi$ boost due to the beam crossing
angle.}.
The acceptance
cuts of formulae (\ref{A1})-(\ref{A4}) are then applied, where the
photon related variables are taken from the DC missing momentum
(photon is not explicitly measured).

\section{EVENT ANALYSIS}

The $\pi^+ \pi^- \gamma$ cross section measurement
contains the following terms:
\begin{equation}
\frac{d\sigma_{hadrons+\gamma}}{dQ^2}=
\frac{dN_{Obs}-dN_{Bkg}}{dQ^2} \cdot 
\frac{1}{\epsilon_{Sel}\epsilon_{Acc}} \cdot
\frac{1}{\int\mathcal{L}dt}
\end{equation}

It requires the study of the various background channels 
($N^{Bkg}$, see following subchapter), the selection 
efficiencies ($\epsilon_{Sel}$) and the systematic effects 
due to the acceptance cuts applied ($\epsilon_{Acc}$).
Finally the counting rate measurement has to be normalized to the
integrated luminosity $\int\mathcal{L}dt$
in order to achieve a cross section 
measurement. All these terms will be obtained from data.
\\
Preliminary results concerning $\epsilon_{Sel}$ are
presented in the following subchapter. 
The detector smearing and the systematic effects, arising from there
have been studied in detail from MC \cite{ich}.
No limititations have been found for a high precision measurement 
on the percent level.
However, more studies - including data - have to be
done. We will also report on the KLOE 
luminosity measurement (Bhabha scattering at large angles). 

\subsection{Background}

{\bf $e^+ e^- \gamma$, $\mu^+ \mu^- \gamma$}\\
Radiative Bhabha events are mostly suppressed by the 
$M_{Track}$ cut (chapter 2.2 (ii)).
The remaining background due to 
events with a high value for $M_{Track}$ and due to 
electrons, which had not been 
rejected by the likelihood method, is peaked at large
$Q^2$ (above $0.7 GeV^2$) and corresponds to a 
contamination below the percent level in this $Q^2$ region. 
\\
Events of the kind $\mu^+ \mu^- \gamma$ 
are not efficiently rejected by the likelihood method, because they
release energy in the EmC with a similar signature like pions. 
After the cuts of formulae (\ref{A1}) - (\ref{A4}) and after 
the $M_{Track}$ cut, the remaining $\mu^+ \mu^-$ cross section 
is low ($\approx 10^{-2} nb$), such that 
we expect only a small contamination ($<1\%$ in the high
$Q^2$ region $> 0.7 GeV^2$).
\\
\\
{\bf $\pi^+ \pi^- \pi^0$ }\\
An important background for our signal is the decay  
$\phi \to \pi^+ \pi^- \pi^0$ (B.R. $15.5 \%$, 
$\sigma_{\pi^+ \pi^- \pi^0}^{tot} \approx 500nb$). 
$\pi^+ \pi^- \pi^0$ and $\pi^+ \pi^- \gamma$ events
are separated in the KLOE standard reconstruction scheme
by a cut in the 2-dimensional plane $M_{Track} - Q^2$. 
At small $Q^2$, the $M_{Track}$ values for the 2 channels 
are very similar and a part of the $\pi^+ \pi^- \pi^0$ events 
appear as a background.
The $M_{Track}$ cut and the acceptance cut of formula 
(\ref{A4}) \footnote{The pion tracks have on the average a lower
momentum as $\pi^+ \pi^- \gamma$ events.} 
reject a big part of these events.
\\
In order to estimate from data the remaining contamination,
we modified the standard cut in the $M_{Track} - Q^2$ plane
by expanding the $\pi^+ \pi^- \pi^0$ selection 
region. We see then the tail of 
$\pi^+ \pi^- \pi^0$ events entering the $M_{Track}$ 
selection interval. We perform this study in bins of
$Q^2$. The $\pi^+ \pi^- \pi^0$ background is 
negligible in most of the $Q^2$ region and gives 
only a contamination at the lower end of the spectrum between
$0.3 GeV^2$ and $0.4 GeV^2$. The effective 
$\pi^+ \pi^- \pi^0$ cross section after all the selection steps
is $< 0.01 nb$. It increases considerably if we select 
$\pi^+ \pi^- \gamma$ events at larger polar angles of the 
photon\footnote{This behaviour can
be easily explained by the missing momentum
of $\pi^+ \pi^- \pi^0$ events (in this case the $\pi^0$),
which is peaked at large angles.}

\subsection{Selection Efficiencies}

We shortly summarize the various efficiencies 
which contribute to the total selection efficiency:

\begin{figure*}[t]
\centering
\includegraphics*[width=90mm]{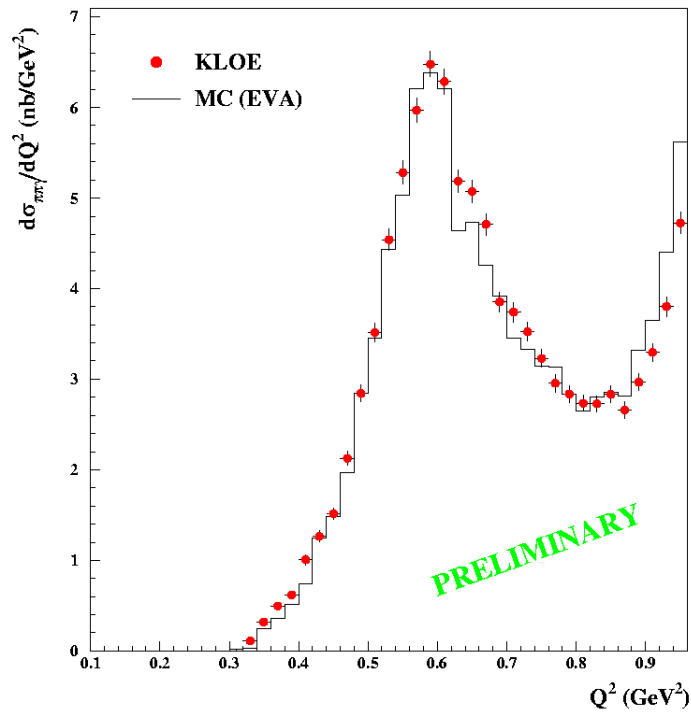}
\caption{The differential $\pi^+ \pi^- \gamma$ 
cross section as a function of the pion invariant mass.
The solid line is the prediction of the theoretical 
event generator EVA of Ref. \cite{binner}.} \label{dsigmadq2}
\end{figure*}

\begin{itemize}

\item{ Trigger: The trigger efficiency has 2 contributions:
the probability of a $\pi^+ \pi^- \gamma$ event to
be recognized by the KLOE trigger 
(between $95\%$ to $99\%$ depending on $Q^2$)
and the inefficiency which arises
from a trigger hardware veto for the filtering of cosmic 
ray events. The second contribution causes an inefficiency 
for $\pi^+ \pi^- \gamma$ events at large
$Q^2$ ($30\%$ at $1GeV^2$) which decreases with lower $Q^2$
(fully efficient at $\approx 0.7GeV^2$).
These values have been obtained from data by looking at the 
individual probabilities for $\pi^+$ and $\pi^-$ 
to fire 1 trigger sector and 1 cosmic veto sector.}

\item{ Reconstruction Filter: A software filter is implemented 
in the KLOE reconstruction program for the filtering of 
non-collider physics 
events, like e.g. machine background and cosmic ray events. 
The inefficiency for $\pi^+ \pi^- \gamma$ events 
caused by this filter is $\approx 2\%$ (taken from MC). }

\item{ Event Selection (see subchapter 2.2): 
The DC vertex efficiency ($\approx 95\%$) is obtained from
the Bhabha stream, which is selected without requiring DC information.
The efficiency due to the likelihood selection 
is $\approx 98\%$ and is evaluated 
from data during the construction of the likelihood method.
The efficiency due to the $M_{Track}$ cut 
($\approx 90\%$) is evaluated from Monte Carlo at present.  }

\end{itemize}

\subsection{Luminosity Measurement}

The DA$\Phi$NE accelerator does not have luminosity monitors 
at small angles (like e.g. LEP) due to the existence of 
focusing quadrupole magnets very close to the interaction point. 
The luminosity is therefore measured using
large angle Bhabha (LAB) events, for which the KLOE detector 
itself can be used. The effective Bhabha cross section 
at large angles 
($55^o < \Theta_{+,-} < 125^o$) is still as high as $425nb$.  
The number of LAB candidates $N_{LAB}$ are counted and normalized to the 
effective Bhabha cross section, obtained from Monte Carlo: 

\begin{equation}
\int\mathcal{L}dt = 
\frac{N_{LAB}(\Theta_i)}{\sigma_{LAB}^{MC}(\Theta_i)}
\cdot (1-\delta_{Bkg})
\end{equation}

Hence, the precision of this measurement depends on:
(i) the theoretical knowledge of the Bhabha scattering process 
including radiative corrections;
(ii) the simulation of the process by the detector simulation
program.
\\

For the theory part we are using 2 independent 
Bhabha event generators 
(the Berends/Kleiss \cite{berends} generator, modified for 
DA$\Phi$NE in \cite{drago} and BABAYAGA \cite{babayaga}). 
\\
We use a selection algorithm for LAB events with a reduced number of 
cuts, for which we expect a very good description by the 
KLOE detector simulation program. The acceptance region for the 
electron and positron polar angle ($55^o < \Theta_{+,-} < 125^o$) is
measured by the EmC clusters produced by these tracks, 
while the energy measurement 
($E_{+,-} > 400 MeV$) is performed by the high resolution drift
chamber. Taking the actual detector resolutions, we expect
the systematic errors arising from these cuts to be well 
below $1\%$. Moreover, the background 
from $\mu^+ \mu^- (\gamma)$, $\pi^+ \pi^- (\gamma)$ and 
$\pi^+ \pi^- \pi^0$ is cut to a level below $1\%$ and
can be easily subtracted. All the selection efficiencies 
concerning the LAB measurement 
(Trigger, EmC cluster, DC tracking) are above $98\%$ and are
well reproduced by the detector simulation. 
\\
As a goal we expect to measure the DA$\Phi$NE luminosity 
at the level of $1\%$. The very good agreement of the 
experimental distributions ($\Theta_{+,-}$, $E_{+,-}$) with
the existing event generators and a cross check with an independent
luminosity counter based on 
$e^+ e^- \to \gamma \gamma (\gamma)$, indicate
a good precision. However, more systematic checks (e.g. 
the effect of a varying beam energy and of a non centered
beam interaction point) are still to be done.

\subsection{Comparison with Monte Carlo}

We analyzed a data sample of $16.4pb^{-1}$
and present in figure \ref{dsigmadq2} the 
preliminary result for the differential cross section 
$d\sigma(e^+ e^- \to \pi^+ \pi^- \gamma) / dQ^2$. 
The plot shows the {\it effective} cross section
after all acceptance cuts ( formulae (\ref{A1}) to (\ref{A4}) ). 
The solid line is the prediction of our $\pi^+ \pi^- \gamma$ 
event generator (Ref. \cite{binner}, called EVA) 
after the detector simulation and after the correction 
for the various selection efficiencies (see subchapter 2.2). 
Up to $Q^2 \approx 0.9 GeV^2$ we obtain an overall good 
agreement between data and MC.
The deviation for $Q^2 > 0.9 GeV^2$ is due to 
a systematic effect connected with the definition of the
fiducial volume for $\Theta_\gamma$. This deviation
will disappear by moving the lower border of the fiducial 
volume from $5^o$ to $0^o$ (not possible with the actual 
version of the event generator; see summary chapter).
\\
The statistical error of the data points in the $\rho$ peak
region is $\approx 2\%$.
The actual version of the event generator has a 
systematic uncertainty of the same size.
By comparing the data and MC distributions we 
conclude that the accuracy of our cross section
measurement is on the level of a few percent, which will be
considerably improved with the ongoing analysis.

\section{SUMMARY AND OUTLOOK}

We presented in this paper the measurement 
$d\sigma(e^+ e^- \to \pi^+ \pi^- \gamma) / dQ^2$ 
for $Q^2 < 1GeV^2$
using the radiative return method. The data sample\footnote{
This corresponds to about one half of the full
data set which KLOE has taken in 2000.} ($16.4 pb^{-1}$) 
shows a good agreement with the existing
event generator. We conclude that the experimental 
understanding of efficiencies, background and luminosity 
are well under control. 
We further conclude that the radiative return is a competitive
method to measure hadronic cross sections, while the 
center-of-mass-energy of the accelerator is fixed. 
This could be proved for the first time in a systematic way 
with the results presented in this paper. 
\\
Data has been taken as a by-product of the KLOE $\phi$ 
physics program and no specific runs were necessary to 
perform this measurement. We stress 
the advantage of the radiative return, that systematic
errors, like luminosity and beam energy, enter only once in
this case and do not have to be known for individual energy
points. 
\\
\\
For the future we plan to refine the actual analysis and to 
change the acceptance cut for the photon polar angle from 
$5^o$ to $0^o$. A comparison with MC will be possible 
in this case with the new
next-to-leading-order event generator \cite{german}
and will improve the precision of
the measurement due to a better systematic control of the 
fiducial volume. 
Also from the statistical point of view this modification
will be helpful, since it corresponds to 
an increase of the effective $\pi^+ \pi^- \gamma$ cross
section of almost a factor 4.\\
Moreover we are studying the possibility to enlarge 
the acceptance region for $\Theta_{+,-}$, which increases
the kinematical acceptance of events at low $Q^2$ ($< 0.3GeV^2$).
\\
In order to improve on $a_\mu^{hadr}$ and to be competitive 
with results coming from the CMD-2 experiment in 
Novosibirsk \cite{cmd2}, a final precision for this 
measurement on the percent level is
needed. A statistical error on this level 
is achieved with a total data sample of $\approx 
200pb^{-1}$ which is in reach for the 
months to come\footnote{The DA$\Phi$NE performance at the moment
(June 2001) is $> 1 pb^{-1}$ per day. }.
\\
\\
We are also investigating the possibility to perform 
an {\it inclusive measurement} 
$d\sigma(e^+ e^- \to \pi^+ \pi^- (n \gamma))/dQ^2$ 
without any cut on the number nor on the kinematics of 
photon(s). The radiative corrections for such a 
measurement are calculated with high precision  
($< 1\%$) and will allow to extract the
pion form factor more precisely. 
The background suppression ($\pi^+ \pi^- \pi^0$)
and the understanding of FSR need further experimental 
and theoretical investigations.

\end{document}